\documentclass[12 pt]{article}
\usepackage{epsfig}
\begin{document}

\title{Physics Beyond the Standard Model: Focusing on the Muon Anomaly}
\author{Helder Chavez\footnote{helderch@if.ufrj.br}\\
Instituto de Fisica, Universidade Federal de Rio de Janeiro. \\POBOX 68528, 21945-910, Rio de Janeiro, Brasil.\\ \\
Cristine N. Ferreira \footnote{crisnfer@cefetcampos.br}\\
N\'ucleo de F\'{\i}sica, Centro Federal de Educa\c c\~ao Tecnol\'ogica de Campos, \\
Rua Dr. Siqueira, 273 - Parque Dom Bosco, \\28030-130, Campos dos Goytacazes, RJ, Brazil,\\ \\
Jos\'e A. Helayel-Neto\footnote{helayel@cbpf.br}\\
Centro Brasileiro de Pesquisas F\'{\i}sicas, Rua Dr. Xavier Sigaud
150, Urca \\
22290-180, Rio de Janeiro, RJ, Brazil,}

\date{\today}

\maketitle

\begin{abstract}
We present a model based on the implication of an exceptional $E_{6}$-GUT symmetry for the anomalous
magnetic moment of the muon. We follow a particular chain of  breakings with
Higgses in the $\mathbf{78}$ and $\mathbf{351}$  representations. We analyse the
radiative correction contributions to the muon mass and the effects of the breaking of the so-called
Weinberg symmetry. We also estimate the range of values of the
parameters of our model.
\end{abstract}

\newpage

\section{Introduction}

Among the known leptons, the muon is potentially interesting for several reasons.
First, its relatively long lifetime of 2.2 $\mu$s\ $(c\tau=658.65\ m)$\ makes
it possible to perform precision measurements. Second, it is sensitive to new
sectors of heavy particles and new interactions. In this sense, the muon anomaly has
provided a stringent test for new theories of Particle Physics, since any new
field or particle which couples to the muon must contribute to $a_{\mu}$. 

\noindent
The most recent results reported by the Muon $(g-2)$
Collaboration \cite{1} have triggered a renewal of interest on the theoretical
prediction of the anomalous magnetic moment of the muon (commonly referred to
as the muon anomaly), $a_{\mu}=\frac{g-2}{2}$, in the Standard Model (SM).
This experimental value is claimed \ to show that there remains a
discrepancy with the SM theoretical calculations at the confidence level of
$2.3\sigma$ to $3.3\sigma$ \cite{1}\cite{2}, if the hadronic light-by-light
contribution, $a_{\mu}^{HHO}(LBL)=80(40)\times10^{-11}\cite{3}$, is used instead
of $a_{\mu}^{HHO}(LBL)=136(25)\times 10^{-11}\cite{4}$, as a consequence that
$e^{+}e^{-}$ annihilation data are used to evaluate this contribution against 
hadronic $\tau$ decays data \cite{5}. Among all contributions that yield
corrections to the muon anomaly, the hadronic contributions are less accurate,
due to the hadronic vacuum polarization effects in the diagrams which use data
inputs coming from the $e^{+}e^{-}$ annihilation cross section and the
hadronic $\tau-$ decays. Also it is not clear, at present, whether the value from
$\tau-$ decay data can be improved much further, due to the difficulty in
evaluating more precisely the effect of isospin breaking \cite{5}.

In fact, these measurements have provided the highest accuracy of the validity
of the different theories for strong, weak and electromagnetic interactions
because they have reached a fabulous relative precision of 0.5 parts per
million (ppm) in the determination of $a_{\mu}$. However, if this confidence
level for the muon anomaly remains, it is possible that we are under a window
open for a New Physics at a high energy scale, $\Lambda.$ The study of the muon
anomaly becomes relevant because it is more sensitive to interactions that are
not predicted in the SM but will be possibly  reached at the CERN Large Hadron
Collider (LHC), with $\sqrt{s}=14\,TeV$ .

On the theoretical side, if we take into account the effects of virtual
massive particles in the diagrams contributing to the lepton anomaly, the
ratios between the corrections to the anomalies are of the order
$\Big({\frac{m_{\mu}}{m_{e}}}\big)^{2}\sim4\times10^{4}$ for the muon and
electron, and of the order $\Big({\frac{m_{\tau}}{m_{e}}}\Big)^{2}%
\sim1.2\times10^{7}$ for the tau and electron. The same huge enhancement
factor would also affect the contributions coming from degrees of freedom
beyond the SM, so that the measurement of the $\tau-$ anomaly would represent
the best opportunity to detect new physics. Unfortunately, the very short
lifetime of the $\tau$- lepton which, precisely because of its high mass, can
also decay into hadronic states, makes such a measurement impossible at
present; this is the reason why there is an emphasis on the muon anomaly.

In this case, it becomes interesting to estimate the order of the correction
of $a_{\mu}$\ in the context of theories beyond the SM. This is done in terms
of powers of $\frac{m_{\mu}}{\Lambda}$. This is related \cite{6} to the
validity or the breaking \smallskip of the chiral symmetry for leptons
together with the\smallskip\ change of sign for $m_{\mu}$. If this symmetry,
which is referred to as Weinberg Symmetry (WS), is respected, then $\Delta
a_{\mu}\sim\left(  m_{\mu}/\Lambda\right)  ^{2}$; on the other hand, if it
is broken, $\Delta a_{\mu}\sim m_{\mu}/\Lambda$\ . \smallskip This is
important because in the latter case the explanation of the muon anomaly may
be given by a new physics at a relatively high energy, whereas in the
former it should appear at a scale close to the electroweak (EW) one.

We consider the 78 and 351 Higgs representations of THE $E_{6}$ 
Grand-Unified Theory (GUT). The representations between square brackets
refer to the $E_{6}$-group, those between brackets refer to 
$SO(10)\otimes\bar U(1)$ and the ones between parentheses correspond to the
$SU(5)\otimes\tilde U(1)$ group. The symmetry breaking pattern
\cite{Robinett1,Robinett2,Helder01,Helder02} is depicted below.
\[%
\begin{array}
[c]{c}%
E_{6}\\
\lbrack\mathbf{78]}\left\{  \mathbf{1,0}\right\} \\
\downarrow
\end{array}
\]%
\[%
\begin{array}
[c]{c}%
SO(10)\otimes\overline{U}(1)\\
\lbrack\mathbf{351]}\left\{  \mathbf{1,-8}\right\} \\
\downarrow
\end{array}
\]%
\[%
\begin{array}
[c]{c}%
SO(10)\\
\lbrack\mathbf{78]}\left\{  \mathbf{45,0}\right\}  (\mathbf{1,0)}\\
\downarrow
\end{array}
\]%
\begin{equation}%
\begin{array}
[c]{c}%
SU(5)\otimes\widetilde{U}(1)\\
\lbrack\mathbf{351]}\left\{  \mathbf{16,-5}\right\}  (\mathbf{1,-5)}\\
\downarrow
\end{array}
\label{1}%
\end{equation}%
\[%
\begin{array}
[c]{c}%
SU(5)\\%
\begin{array}{ll}
{\lbrack\mathbf{351]}\left\{  \mathbf{54,4}\right\}
(\mathbf{24,0),}} \\
{[\mathbf{351]}\left\{  \mathbf{144,1}\right\}
(\mathbf{24,5)}}
\end{array}
\\
\downarrow
\end{array}
\]%
\[%
\begin{array}
[c]{c}%
SU(3)_{C}\otimes SU(2)_{L}\otimes U(1)\\
\lbrack\mathbf{351]}\left\{  \mathbf{10,-2}\right\} \\
\downarrow
\end{array}
\]%
\[
SU(3)_{C}\otimes U(1)_{e.m}%
\]
The order of magnitude of the contribution is $\Delta a_{\mu}\sim m_{\mu}/m_{M}$, where
$m_{M}$ is the mass of the exotic fermion. This fermion is analogous to the ordinary
muon contained in the $[\mathbf{27}]$ representation of fermions in
$\{\mathbf{10},-2\}$ under $SO(10)\times\bar{U}(1)$. This connection makes sense if
the radiative correction to the muon mass is small and if there occurs breaking of WS.
On the other hand, if the muon mass is only due to radiative
corrections, the right mixing angle between leptons is zero and  WS is not broken.

Our paper is organized as follows. In the Section 2, we discuss the WS in the
SM in connection with the order of magnitude of the muon anomaly. In the
Section 3, we present our model, considering the sequences of breakings of
symmetries (\ref{1}). In Section 4, we analyse the question of the radiative mass of
the muon due to the mixings with the massive fermion that occur in the breaking
chain $SU(5)\longrightarrow SU(3)_{C}\otimes SU(2)_{L}\otimes U(1)$ with
$\left\{  \mathbf{144,1}\right\}  $ Higgs; in Section 5, we analyse WS in the
context of our model and, finally, in Section 6, we present our General Conclusions.

\section{\textbf{WS and the anomalous magnetic moment in the SM}}

The WS is a well-known property \cite{6} of the SM of Particle Physics. In this
section, we briefly review its main points, since this result is connected with
the order of magnitude of the $\Delta a_{\mu}$ contribution in the $E_{6}$
model. The mass term $m_{\mu}\overline{\mu}\mu$ breaks chiral symmetry; the
field redefinition below changes the sign of the mass term:%

\begin{equation}
{\Large \mu\rightarrow\gamma}_{5}{\Large \mu\quad,\quad m}_{\mu}%
{\Large \rightarrow-m}_{\mu}{\Large \, ,} \label{2}%
\end{equation}
where $\mu$ is the field variable associated to the muon.

If the WS Eq. \smallskip(\ref{2}) is valid, the corrections to $a_{\mu}$\ must be of
even powers of the ratio of $m_{\mu}$\ to a larger scale $\Lambda:$%
\begin{equation}
{\Large a}_{\mu}{\Large =c}_{o}\left(  \frac{m_{\mu}}{\Lambda}\right)
^{0}{\Large +c}_{2}\left(  \frac{m_{\mu}}{\Lambda}\right)  ^{2}%
{\Large +...\quad.} \label{3}%
\end{equation}
The effective interaction that gives a non-zero contribution to the muon
anomalous magnetic moment is $a_{\mu}\frac{e}{4m_{\mu}}\overline{\mu}%
\sigma_{_{\alpha\beta}}\mu F^{^{\alpha\beta}}$; for the SM version, it may be
written as
\begin{equation}
\mathcal{L}_{\mathbf{eff}}\ {\Large =a}_{\mu}{\Large \,}\frac{e}{4m_{\mu}%
}\left(  \overline{\Psi}_{L}\sigma^{\alpha\beta}\mu_{R}\frac{f_{0}%
\,\varphi_{V}}{m_{\mu}}+h.c.\right)  {\Large F}_{\alpha\beta}{\Large \quad,}
\label{4}%
\end{equation}
with a Higgs field doublet $\varphi=\left(
\begin{array}
[c]{c}%
0\\
\varphi_{1}%
\end{array}
\right)  =\varphi_{V}+\left(
\begin{array}
[c]{c}%
0\\
h_{1}/\sqrt{2}%
\end{array}
\right)  $ , such that
\begin{equation}
\overline{\Psi}_{L}\ {\Large =}\left(
\begin{array}
[c]{c}%
\nu\\
\mu
\end{array}
\right)  _{L}{\Large \quad,\quad\varphi}_{V}\ {\Large =}\left(
\begin{array}
[c]{c}%
0\\
\frac{\upsilon_{1}}{\sqrt{2}}%
\end{array}
\right)  {\Large \quad,\,\,}f_{0}\frac{\upsilon_{1}}{\sqrt{2}}\ {\Large =\ }%
m_{\mu}\,{\Large \,.} \label{5}%
\end{equation}
Now, to have the WS invariance (2) in the SM, one must perform the transformations
\begin{equation}
\Psi_{L}{\Large \rightarrow\gamma}_{5}{\Large \,}\Psi_{L}{\Large =-}\Psi
_{L}{\Large \quad,\quad\mu}_{R}{\Large \rightarrow\gamma}_{5}{\Large \mu}%
_{R}{\Large =\mu}_{R}{\Large \quad,\quad\varphi\rightarrow-\varphi\ .} \label{6}%
\end{equation}

We can prove that the neutral current Lagrangian density reads as

\begin{equation}
\mathcal{L}_{NC}=-e\overline{\mu}{\Large \gamma}^{\alpha}{\Large \mu
A}_{\alpha}-\frac{g}{2\cos\theta_{W}}\overline{\mu}{\Large \gamma}^{\alpha
}\left(  \upsilon_{z}-a_{z}\gamma^{5}\right)  {\Large \mu Z}_{\alpha};
\label{7}
\end{equation}
the charged current Lagrangian density is written as
\begin{equation}
\mathcal{L}_{CC}=\frac{g}{2\sqrt{2}}\left[  \overline{\nu}_{\mu}%
{\Large \gamma}^{\alpha}\left(  1-\gamma^{5}\right)  {\Large \mu W}_{\alpha
}^{\left(  +\right)  }+\overline{\mu}{\Large \gamma}^{\alpha}\left(
1-\gamma^{5}\right)  {\Large \nu}_{\mu}{\Large W}_{\alpha}^{\left(  -\right)
}\right]  {\Large \ ,} \label{8}%
\end{equation}
and the Yukawa sector
\begin{equation}
\mathcal{L}_{YUK}=-f_{0}\left(  \overline{\mu}_{R}\varphi^{\dagger}\mu
_{L}+\overline{\mu}_{L}\varphi\mu_{R}\right)  =-\frac{1}{\sqrt{2}}f_{0}\left(
\upsilon_{1}+h_{1}\right)  \overline{\mu}\mu , \label{9}%
\end{equation}
where $m_{\mu}=f_{0}\frac{\upsilon_{1}}{\sqrt{2}}$ is the muon mass and the
interactions are invariant under the transformations  of Eq.(\ref{6}). Therefore, the
corrections to $a_{\mu}$ are of the type of Eq.(\ref{3}) with the EW scale,
$\Lambda$. The first term is the electromagnetic contribution $c_{0}%
=\frac{\alpha}{2\pi}+...$, computed recently up to $(\alpha/\pi)^{5}$ \cite{7};
 the second term, $c_{2}\left(  \frac{m_{\mu}}{\Lambda}\right)  ^{2}\sim
a_{\mu}^{QED}\times1,7\times10^{-6}\simeq2\times10^{-9}$, corresponds to the
weak contribution.

\section{An \textbf{\ alternative E}$_{\mathbf{6}}$-model\textbf{\ for the muon
anomaly}}

\bigskip The exceptional group $E_{6}$ \cite{8} was proposed as an alternative
to $SU(5)-$and $SO(10)-$models, and it is actually, in many 
aspects, the preferred gauge group 
for Grand Unification. In this section, let us discuss the pattern of
breakings (\ref{1}) based on the $[\mathbf{78]}$ and $[\mathbf{351]}$
representations. The ordinary fermions of the SM are contained in the
$\left\{  \mathbf{16},1\right\}  \subset$ $\mathbf{27-}$ dimensional
representation:
\begin{equation}
\lbrack\mathbf{27]=}\left\{  \mathbf{16},1\right\}  \oplus\left\{
\mathbf{10},-2\right\}  \oplus\left\{  \mathbf{1},4\right\}  . \label{10}%
\end{equation}
There are $11$ additional fermions with respect to the SM fermions.
For the first generation, these particles are:
\begin{equation}%
{\underbrace{\mathbf{\Psi}_{L}}_{\left\{  \mathbf{1}%
,4\right\}  }}%
\oplus%
{\underbrace{{\underbrace{\left(
\mathbf{D}^{C}\quad\mathbf{N\quad E}\right)  _{L}}_{\left(  \overline
{\mathbf{5}},-2\right)  }}\oplus{\underbrace{\left(
\mathbf{D}\quad\mathbf{N}^{C}\mathbf{\quad E}^{C}\right)  _{L}}_{\left(
\mathbf{5},2\right)  }}}_{\left\{  \mathbf{10},-2\right\}  }}%
. \label{11}%
\end{equation}
The gauge bosons are contained in the adjoint $\mathbf{78-}$dimensional
representation, that, with respect to $SO(10)\otimes\overline{U}(1)$, is
decomposed as below:
\begin{equation}
\lbrack\mathbf{78]=}\left\{  \mathbf{45},0\right\}  \oplus\left\{
\mathbf{16},-3\right\}  \oplus\left\{  \mathbf{1},0\right\}  \oplus\left\{
\overline{\mathbf{16}},3\right\}  . \label{12}%
\end{equation}

For the first generation, the exotic fermions of the $\mathbf{10}$
representation of $SO(10)$ can acquire mass from the Higgs $\{\mathbf{54},4\}
$ of $\ $the $[\mathbf{351]}$ representation of $E_{6}$, because
$\{\mathbf{10\}\otimes}\{\mathbf{10\}=\{54\}\oplus\{45\}\oplus\{1\}}$ . The
mass terms are of the type \cite{9} $\ $%
\begin{equation}
\varphi_{2}\left(  \mathbf{54,24}\right)  \left(  D^{c}D-\frac{3}{2}%
E^{c}E-\frac{3}{2}N^{c}N\right)  . \label{13}%
\end{equation}
In this same representation, $\{\mathbf{144,}1\}$, let us mix these fermions
with the ordinary ones, because both components contain a $\mathbf{24}$ of
$SU(5)$, which has one invariant component under $SU(3)_{C}\otimes
SU(2)_{L}\otimes U(1)_{Y}$ : $\{\mathbf{16\}\otimes\{10\}=\{144\}\oplus
\{}\overline{\mathbf{16}}\}.$ This mixing term is given by
\begin{equation}
\varphi_{3}\left(  \mathbf{144,24}\right)  \left(  d^{c}D-\frac{3}{2}%
E^{c}e-\frac{3}{2}N^{c}\nu\right)  . \label{14}%
\end{equation}
Observe that both Higges, $\varphi_{2}\left(  \mathbf{54,24}\right)  $ and
$\varphi_{3}\left(  \mathbf{144,24}\right)$, being singlets $\left(
\mathbf{1,1,}0\right)  $ under $SU(3)_{C}\otimes SU(2)_{L}\otimes U(1)_{Y}$,
we shall assume that they take \ diferent values of expectation around his
quantum fields $h_{2}$ and $h_{3}$:
\begin{eqnarray}
\varphi_{2}\left(  \mathbf{54,24}\right)   &  =\frac{1}{\sqrt{2}}\left(
\upsilon_{2}+h_{2}\right) \label{15}\\
\varphi_{3}\left(  \mathbf{144,24}\right)   &  =\frac{1}{\sqrt{2}}\left(
\upsilon_{3}+h_{3}\right),  \label{16}%
\end{eqnarray}
where the v.e.v's $\upsilon_{3}$ and $\upsilon_{2}$ we will assume them
to satisfy the relation $\upsilon_{3}\leq\upsilon_{2}.$

On the other hand, the ordinary fermions of the SM get masses from the Higgs
$\left\{  \mathbf{10},-2\right\}  $, because the Yukawa term that conserves the
$\overline{U}(1)$ charge is
\begin{equation}
\{\mathbf{16\}\otimes}\{\mathbf{16\}=}\{\mathbf{10\}\oplus}%
\{\mathbf{126\}\oplus}\{\mathbf{120\}} , \label{17}%
\end{equation}
and this Higgs is in the $[\mathbf{351].}$ This mass term is
\begin{equation}
H\left(  \mathbf{10,}\overline{\mathbf{5}}\right)  \left(  d^{C}d+e^{C}%
e+N^{C}L\right)  . \label{18}%
\end{equation}
In order to explain the notation, here $\varphi\,%
\acute{}%
\left(  \mathbf{a,24}\right)  $ stands for the component of the Higgs
representation, $\varphi\,%
\acute{}%
,$ where the label $\mathbf{a}$ indicates the transformation under $SO(10)$
and the label $\mathbf{24}$-component refers to $SU(5)$; similarly, for
$H\left(  \mathbf{10,}\overline{\mathbf{5}}\right)  .$ In fact, this Higgs
$H\left(  \mathbf{10,}\overline{\mathbf{5}}\right)  $ is indeed that one of
the SM $\varphi_{1}\left(  \mathbf{1,2,}1/2\right)  $ under $SU(3)_{C}\otimes
SU(2)_{L}\otimes U(1)_{Y}$ which is, as we already said before, written as

\begin{equation}
\varphi_{1}\left(  \mathbf{1,2,}1/2\right)  =\frac{1}{\sqrt{2}}\left(
\upsilon_{1}+h_{1}\right)  . \label{19}%
\end{equation}

Now, let us extend this for the second generation of fermions, and call $M$
the supermassive fermion in analogy to the ordinary muon of the SM.

If the breakings of symmetry are due to a
$[\mathbf{351]}$\ , when the GUT symmetry is broken, the mass eigenstates
$\mu_{o}$\ \smallskip and $\widehat{M}$\ are determined by the
expectation\smallskip\ values of the $\left(  SO(10),SU(5)\right)
$\ multiplets\smallskip\ $\varphi_{2}\,\left(  \mathbf{54,24}\right)  $\ and
$\varphi_{3}\left(  \mathbf{144,24}\right)  $, through the mixture of left and
right components \cite{9}\cite{10}:
\begin{equation}
\left(
\begin{array}
[c]{c}%
{\Large \mu}_{L,R}\\
M_{L,R}%
\end{array}
\right)  =\left(
\begin{array}
[c]{cc}%
\cos{\Large \theta}_{L,R} & \sin{\Large \theta}_{L,R}\\
-\sin{\Large \theta}_{L,R} & \cos{\Large \theta}_{L,R}%
\end{array}
\right)  \left(
\begin{array}
[c]{c}%
{\Large \mu}_{L,R}^{0}\\
\widehat{M}_{L,R}%
\end{array}
\right),  \label{20}%
\end{equation}
where $\theta_{L,R}$ are the left nd right mixing angles, respectively. It is
possible that the mixing angle $\theta_{R}$ is small, of the order $\sim
m_{\mu}/m_{M}$, where $m_{M}$ is the mass of the heavy muon, $M$, however,
due to the weak universality, the angle $\theta_{L}$ between $\mu_{L}$ and
$M_{L}$ is expected to be the same mixing angle for \smallskip\ $\nu^{\mu
}$\ and the neutral exotic lepton $N$; but $\theta_{L}$ can still be large
\cite{11}.

The fermion-Higgs interaction Lagrangian is given by:
\begin{eqnarray}
L=\frac{f_{0}}{\sqrt{2}}\overline{\mu_{L}}\mu_{R}\,(h_{1}+\upsilon_{1}%
)+\frac{f_{1}}{\sqrt{2}}\overline{M_{L}}M_{R}\,(h_{2}+\upsilon_{2}%
)+\frac{f_{2}}{\sqrt{2}}\overline{\mu_{L}}M_{R}\,(h_{3}+\upsilon
_{3})+\nonumber\\
\frac{f_{3}}{\sqrt{2}}\overline{M_{L}}\mu_{R}\,(h_{1}+\upsilon_{1})+h.c. ,
\label{21}%
\end{eqnarray}
where some of the $\ f_{i}%
\acute{}%
\,$s could be vanishing. The previous expression can be written as below:
\begin{equation}
L=\left(
\begin{array}
[c]{cc}%
\overline{\mu_{L}} & \overline{M_{L}}%
\end{array}
\right)  \frac{1}{\sqrt{2}}\left(
\begin{array}
[c]{cc}%
f_{0}\upsilon_{1} & f_{2}\upsilon_{3}\\
f_{3}\upsilon_{1} & f_{1}\upsilon_{2}%
\end{array}
\right)  \left(
\begin{array}
[c]{c}%
\mu_{R}\\
M_{R}%
\end{array}
\right)  . \label{22}%
\end{equation}
The mass matrix reads as:%

\begin{equation}
\mathbf{M=}\frac{1}{\sqrt{2}}\left(
\begin{array}
[c]{cc}%
f_{0}\upsilon_{1} & f_{2}\upsilon_{3}\\
f_{3}\upsilon_{1} & f_{1}\upsilon_{2}%
\end{array}
\right)  . \label{23}%
\end{equation}
As usually, the previous matrix mass is diagonalized by a bi-unitary
transformation \cite{10} \cite{12} $U_{L}^{\dagger}\mathbf{M}U_{R}%
=\mathbf{M}_{diag}$, where $U_{L,R}$ is given in (\ref{20}). From $U_{L}^{\dagger
}\mathbf{MM}^{\dagger}U_{L}=\mathbf{M}_{diag}^{2}\ ,$ it is possible to find
\begin{equation}
\tan\left(  2\theta_{L}\right)  =\frac{2\left(  f_{0}f_{3}\upsilon_{1}%
^{2}+f_{1}f_{2}\upsilon_{2}\upsilon_{3}\right)  }{\left(  f\ _{3}^{2}%
-f\ _{0}^{2}\right)  \upsilon_{1}^{2}+f\ _{1}^{2}\upsilon_{2}^{2}-f\ _{2}%
^{2}\upsilon_{3}^{2}}\ ; \label{24}%
\end{equation}
on the other hand, from $U_{R}^{\dagger}\mathbf{M}^{\dagger}\mathbf{M}%
U_{R}=\mathbf{M}_{diag}^{2}$ , we obtain
\begin{equation}
\tan\left(  2\theta_{R}\right)  =\frac{2\left(  f_{0}f_{2}\upsilon_{1}%
\upsilon_{3}+f_{1}f_{3}\upsilon_{1}\upsilon_{2}\right)  }{f\ _{1}^{2}%
\upsilon_{2}^{2}+f\ _{2}^{2}\upsilon_{3}^{2}-\left(  f\ _{3}^{2}+f\ _{0}%
^{2}\right)  \upsilon_{1}^{2}}\ . \label{25}%
\end{equation}
In the limit for which all the couplings $f_{i}$ are equal and 
$\upsilon_{3}
\simeq\upsilon_{2}\gg\upsilon_{1}$, we heve to $\tan\left(  2\theta_{L}\right)  \rightarrow\infty $,
 
\begin{equation}
\tan\left(  2\theta_{R}\right)
\simeq\frac{2\upsilon_{1}\upsilon_{2}}{\upsilon_{2}^{2}-\upsilon_{1}^{2}
}
\end{equation}  
or to the algles $ \theta
_{L}$ and  $\theta_{R}$ the values $ \theta
_{L}=\frac{\pi}{4}$ , $\theta_{R}\sim\frac{\upsilon_{1}}{\upsilon_{2}}$.
As it
can be seen, in this case $\theta_{R}$ is small.

The part of the interaction Lagrangian for the quantum flutuations can be written as:%
\begin{equation}
L=\frac{f_{0}}{\sqrt{2}}\overline{\mu_{L}}\mu_{R}\,h_{1}+\frac{f_{1}}{\sqrt
{2}}\overline{M_{L}}M_{R}\,h_{2}+\frac{f_{2}}{\sqrt{2}}\overline{\mu_{L}}%
M_{R}\,h_{3}+\frac{f_{3}}{\sqrt{2}}\overline{M_{L}}\mu_{R}\,h_{1}+h.c.;
\label{26}%
\end{equation}
after the mixing equations (\ref{20}), we obtain the changing-flavor
Lagrangian:

\begin{eqnarray}
L_{FC}=\frac{f_{0}}{\sqrt{2}}(c_{L}s_{R}\overline{\mu_{L}^{0}}\widehat{M}%
_{R}+c_{R}s_{L}\overline{\widehat{M}_{L}}\,\mu_{R}^{0})h_{1}+\frac{f_{1}%
}{\sqrt{2}}(-s_{L}c_{R}\overline{\mu_{L}^{0}}\widehat{M}_{R}-c_{L}%
s_{R}\overline{\widehat{M}_{L}}\,\mu_{R}^{0})\,h_{2}+\label{27}\\
\frac{f_{2}}{\sqrt{2}}(c_{R}c_{L}\overline{\mu_{L}^{0}}\widehat{M}_{R}%
-s_{L}s_{R}\,\overline{\widehat{M}_{L}}\,\mu_{R}^{0})h_{3}+\frac{f_{3}}%
{\sqrt{2}}(-s_{L}s_{R}\overline{\mu_{L}^{0}}\widehat{M}_{R}+c_{L}%
c_{R}\overline{\widehat{M}_{L}}\,\mu_{R}^{0})\,h_{1}+h.c.\nonumber
\end{eqnarray}
where $c_{L,R}=\cos\theta_{L,R}$ and $s_{L,R}=\sin\theta_{L,R}$ . We label the
neutral mass eigenstates  of the Higgses  by $H_{1},H_{2},H_{3}$
whose masses are $M_{1},M_{2},M_{3}$ respectively. Then, suitable rotations between
of fields $h_{1},h_{2},h_{3}$ must diagonalize the mass matrix in the
potential $V(h_{1},h_{2},h_{3}).$ We suppose (from now in ahead) that
$M_{1}\ll M_{3}\simeq M_{2},$ assuming conservation of CP, the matrix of
rotations will be real. In the limit $\upsilon_{1}\ll\upsilon_{3}%
\leq\upsilon_{2}$, the state $h_{1}\longrightarrow H_{1}$ is weak and any
appreciable mixing between scalars will only appear between $h_{2}$ and
$h_{3}:$%
\begin{equation}
\left(
\begin{array}
[c]{c}%
h_{2}\\
h_{3}%
\end{array}
\right)  =\left(
\begin{array}
[c]{cc}%
\cos\vartheta & -\sin\vartheta\\
\sin\vartheta & \cos\vartheta
\end{array}
\right)  \left(
\begin{array}
[c]{c}%
H_{2}\\
H_{3}%
\end{array}
\right)  , \label{28}%
\end{equation}
with $\vartheta$ being the angle of rotation that allows the diagonalization
of the matrix. With these mixings of neutral scalars fields, the
flavor-changing Lagrangian (\ref{27}) now takes the form:%
\begin{equation}
L_{FC}^{eff}=\frac{1}{2\sqrt{2}}%
{\displaystyle\sum\limits_{i=1}^{3}}
\overline{\mu^{0}}\left[  \widetilde{\beta}_{i}+\widetilde{\alpha}_{i}%
-\gamma_{5}(\widetilde{\beta}_{i}-\widetilde{\alpha}_{i})\right]  \widehat
{M}H_{i}+h.c., \label{29}%
\end{equation}
where%
\begin{eqnarray}
\widetilde{\alpha}_{1}=f_{0}c_{L}s_{R}-f_{3}s_{L}s_{R},\nonumber\\
\widetilde{\beta}_{1}=f_{0}c_{R}s_{L}+f_{3}c_{L}c_{R},\nonumber\\
\widetilde{\alpha}_{2}=-f_{1}c_{R}s_{L}\cos\vartheta+f_{2}c_{R}%
c_{L}\sin\vartheta\nonumber\\
\widetilde{\beta}_{2}=-f_{1}c_{L}s_{R}\cos\vartheta-f_{2}s_{L}s_{R}%
\sin\vartheta,\label{30}\\
\widetilde{\alpha}_{3}=f_{1}c_{R}s_{L}\sin\vartheta+f_{2}c_{L}c_{R}%
\cos\vartheta,\nonumber\\
\widetilde{\beta}_{3}=f_{1}c_{L}s_{R}\sin\vartheta-f_{2}s_{L}s_{R}%
\cos\vartheta.\nonumber
\end{eqnarray}
%

\begin{figure}
[ptb]
\begin{center}
\includegraphics[
height=1.5454in,
width=3.1981in
]%
{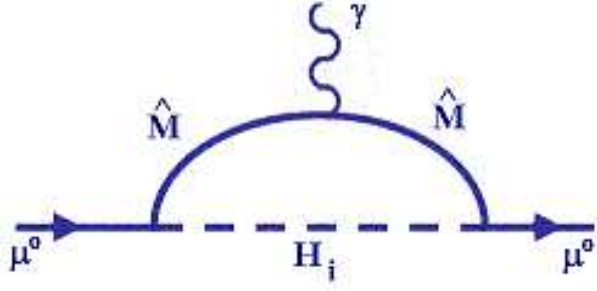}%
\caption{Contributions \ with Higgs-interchange to the muon anomalous magnetic
moment. }%
\end{center}
\end{figure}

The generic diagram with Higgs interchange contributing to the anomaly of the
muon is shown in Fig.1. In fact, the explicit calculation \cite{13} of the
one-loop contribution yielded by Eq. (29) gives the results (in the limit
$m_{M}/m_{\mu}\gg1$) :%

\begin{figure}
[ptb]
\begin{center}
\includegraphics[
height=2.5374in,
width=2.9283in
]%
{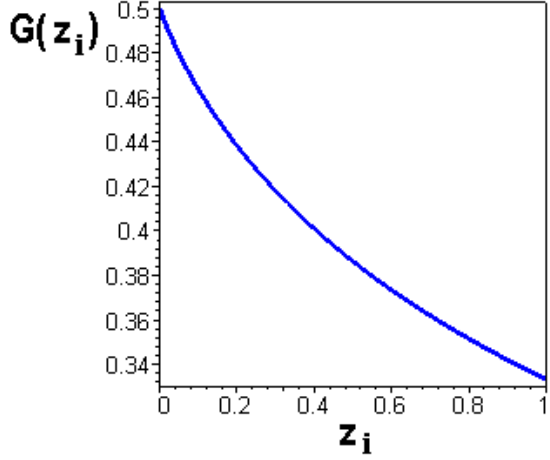}%
\caption{Plot of $G(z_{i})$ as function of $z_{i},$ where $z_{i}=M_{i}%
^{2}/m_{M}^{2}$ $.$}%
\end{center}
\end{figure}
\
\begin{eqnarray}
{\Large \Delta a}_{\mu}^{FCh}\ =\frac{1}{128\pi^{2}}m_{\mu}^{2}\sum_{i=1}%
^{3}\xi_{i}^{2}%
{\displaystyle\int\limits_{0}^{1}}
dx\frac{\left(  x^{2}-x^{3}\right)  +\frac{m_{M}}{m_{\mu}}x^{2}}{m_{\mu}%
^{2}x^{2}+\left(  m_{M}^{2}-m_{\mu}^{2}\right)  x+M_{i}^{2}\left(  1-x\right)
}=\nonumber\\
={\Large \ }\frac{1}{128\pi^{2}}\frac{m_{\mu}}{m_{M}}\sum_{i=1}^{3}\xi_{i}%
^{2}G(z_{i}){\Large \quad,} \label{31}%
\end{eqnarray}
denoting $\ z_{i}=\frac{M_{i}^{2}}{m_{M}^{2}}$ where $\ \xi_{i}^{2}=$%
\ $\alpha_{i}^{2}-\beta_{i}^{2}$ with \ $\alpha_{i}=\widetilde{\beta}%
_{i}+\widetilde{\alpha}_{i}$ , $\beta_{i}=\widetilde{\beta}_{i}-\widetilde
{\alpha}_{i}$ the function $G(z_{i})=\frac{\left(  1-z_{i}\right)
^{2}-2z_{i}\left(  1-z_{i}\right)  -2z_{i}^{2}\ln z_{i}}{2\left(
1-z_{i}\right)  ^{3}}$ is plotted in Fig. (2)$.$ Let us see two cases of interest:

a) $m_{M}\simeq M_{2}\simeq M_{3}\gg M_{1}$ $.$ If we consider the rough case
in that $f_{1}=f_{2}$ , we have $\xi_{3}^{2}=\xi_{2}^{2}$ with the reasonable
value $G\left(  z_{2,3}\right)  \simeq0.3$ and $G\left(  z_{1}\right)
\simeq0.5$ In this case the total contribution is%
\begin{equation}
{\Large \Delta a}_{\mu}^{FCh}\ =\frac{1}{128\pi^{2}}\frac{m_{\mu}}{m_{M}%
}(0.5\times\xi_{1}^{2}+0.6\times\xi_{2}^{2})\quad, \label{32}%
\end{equation}
then, for to complete the anomaly value \cite{2} ${\Large \Delta a}_{\mu
}=25.2\times10^{-10},$ we have%
\begin{equation}
7.4\times10^{-3}\leq\xi_{1}^{2}+1.2\times\xi_{2}^{2}\leq 0.64 \label{33}%
\end{equation}
where we consider $115\,GeV\leq m_{M}\leq 10\,TeV.$%

\begin{figure}
[ptb]
\begin{center}
\includegraphics[
height=2.5374in,
width=2.949in
]%
{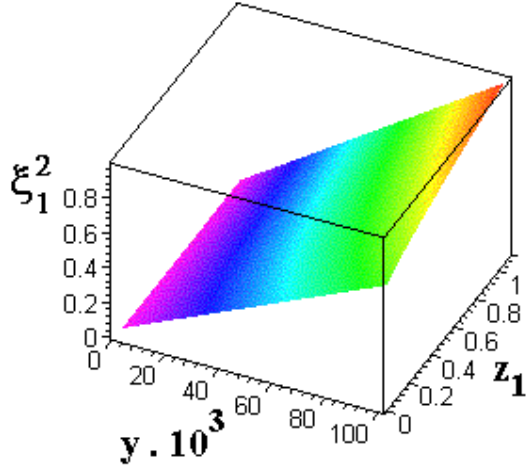}%
\caption{Space of values of $\xi_{1}^{2}$ in the range of masses
$115\,GeV\leq m_{M}\leq 10TeV,$ $115\,GeV\leq M_{1}\leq
700$ GeV for to complate the anomaly value$,$ where $y=\frac{m_{M}%
}{m_{\mu}}.$}%
\end{center}
\end{figure}

b) $M_{2}\simeq M_{3}\gg M_{1}\simeq m_{M}.$ The principal contribution come
from $H_{1}$%
\begin{equation}
{\Large \Delta a}_{\mu}^{FCh}\ (H_{1})\simeq\frac{1}{128\pi^{2}}\frac{m_{\mu}%
}{m_{M}}\xi_{1}^{2}\ G(z_{1})\,, \label{34}%
\end{equation}
and this case $G(z_{2,3})\longrightarrow0.$ We can find the limits of $\xi
_{1}^{2}$ over the range masses indicated $7\times10^{-3}\leq \xi_{1}%
^{2}\ \leq 0.61,$ as illustred in Fig. 3.

\section{Radiative corrections to the muon mass}

Other interesting possibility is to suppose a situation in which the  muon  
mass comes only from radiative corrections. 
There are models of this type \cite{15} \cite{16} in the literature. In
the Ref. \cite{16}, the authors, working out an $SU(3)\otimes SU(3)\otimes
U(1)$ model, introduce some symmetries to avoid \ the light fermions from
acquiring their masses at tree-level through their couplings to the
SM Higgs boson with non-zero vacuum expectation value; as a consequence, the muon
gets its mass from the radiative corrections induced by other particles.

The one-loop correction to the muon mass is obtained by removing the photon
line from the diagram Fig.(1). The amplitude for this diagram is:
\begin{eqnarray}
\Sigma\left(  p\right)   &  =-i\kappa^{2}\Big[\int\frac{d^{4}q}{\left(
2\pi\right)  ^{4}}\frac{m_{M}\left(  \left\vert \alpha_{i}\right\vert
^{2}-\left\vert \beta_{i}\right\vert ^{2}\right)  +\left(  \left\vert
\alpha_{i}\right\vert ^{2}+\left\vert \beta_{i}\right\vert ^{2}\right)
\gamma^{\mu}q_{\mu}}{\left(  q^{2}-m_{M}^{2}\right)  \left(  (p-q)^{2}%
-M_{i}^{2}\right)  } + \nonumber \\ 
& +\int\frac{d^{4}q}{\left(
2\pi\right)  ^{4}} \frac{+(\alpha_{i}\beta_{i}^{\dagger}+\beta_{i}\alpha_{i}^{\dagger})q_{\mu
}\gamma^{\mu}\gamma_{5}+ m_{M}(\alpha_{i}\beta_{i}^{\dagger}-\beta_{i}%
\alpha_{i}^{\dagger}\,)\gamma_{5}}{\left(  q^{2}-m_{M}^{2}\right)  \left(  (p-q)^{2}%
-M_{i}^{2}\right) }\Big], \label{35}%
\end{eqnarray}
where $\kappa=\frac{1}{2\sqrt{2}}$ and\ $\ i=1,2,3.$ Let us suppose that
$M_{2}$ is the maximal energy scale for our model, then, as $m_{\mu}\ll
m_{M},M_{2}$, we obtain the folowing expression for the radiately induced muon mass:
\begin{eqnarray}
m_{\mu}^{\mbox{1-loop}}=\frac{\left(  \alpha_{2}^{2}-\beta_{2}^{2}\right)
}{8\left(  4\pi\right)  ^{2}}m_{M}F\left(  z_{2}\right)  ,\label{36}\\
F\left(  z_{2}\right)  =1-\frac{1}{z_{2}-1}\ln z_{2}, \label{37}%
\end{eqnarray}
where $z_{2}=$ $\frac{M_{2}^{2}}{m_{M}^{2}}.$ Notice that, for $M_{2}\simeq
M_{3}\gg m_{M},M_{1}$ (or $z_{2,3}\gg1)$, the function $F\left(  z_{2}\right)
$ takes it asymptotic value equal to 1, then
\begin{equation}
m_{\mu}^{\mbox{loop}}(H_{2},H_{3})\simeq\frac{\xi_{2}^{2}}{128\pi^{2}}m_{M},
\label{38}%
\end{equation}
and for the case $M_{2}\simeq M_{3}\simeq m_{M}\gg M_{1}$ the function
$F\left(  z_{2}\right)  \simeq1-m_{M}^{2}/M_{2}^{2}.$ To assure  small
radiative mass for the muon, for example of $\ 0.1$ $MeV-10\,MeV$ \ with
$115\,GeV\leq  m_{M}\leq 10\,TeV,$ it is necessary that $1.0\times
10^{-3}\leq\xi_{2}^{2}\leq1.3\times10^{-3}.$%

\begin{figure}
[ptb]
\begin{center}
\includegraphics[
height=1.1588in,
width=2.5659in
]%
{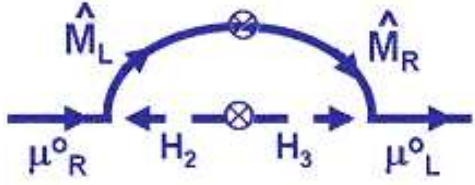}%
\caption{Diagram of radiative correction to muon mass with mixing bettween
heavy scalar. }%
\end{center}
\end{figure}

There is another diagram that can contribute to the radiative mass of the muon,
as it is shown in Fig. 4. The result was estimated \cite{14} \cite{16} as

\begin{equation}
m_{\mu}^{\mbox{1-loop}}\simeq\frac{\varepsilon}{16\pi^{2}}m_{M}\left[
\frac{M_{2}^{2}}{m_{M}^{2}-M_{2}^{2}}\ln\left(  \frac{m_{M}^{2}}{M_{2}^{2}%
}\right)  -\frac{M_{3}^{2}}{m_{M}^{2}-M_{3}^{2}}\ln\left(  \frac{m_{M}^{2}%
}{M_{3}^{2}}\right)  \right]  \,, \label{39}%
\end{equation}
where $\varepsilon$ is an parameter function of Yukawa couplings that (can
read from (29) and (30)) and of the mixing angle $\vartheta.$ However,
$\widehat{M},$ $H_{2}$ and $H_{3}$ for the limit natural $m_{M}\ll M_{2}\simeq
M_{3}$ , $m_{\mu}^{\mbox{1-loop}}$ is essentially zero.

In our model, the ordinary fermions are massles at the tree level in the GUT
scale (i.e no bare $m_{\mu}^{\mbox{o}}$ \ is possible due to symmetry), but it
couples to the heavy fermion $\widehat{M}$ \ through the mixing with scalars, according to the breaking  $SU(5)\otimes\widetilde
{U}(1)\longrightarrow SU(5)$. If we suppose this,
then the only diagrams that contribute to the anomaly are those with the interchange of
$H_{2}$ and $H_{3}$ in the Fig. (1). To simplify, let us suppose the case
$M_{2}\simeq M_{3}$ and $f_{1}=f_{2}$ from which $\xi_{2}^{2}=\xi_{3}^{2}$;
then, the contribution with the $H_{2}$-interchange is%
\begin{equation}
\Delta a_{\mu}^{FCh}=\frac{\xi_{2}^{2}}{128\pi^{2}}\frac{m_{\mu}}{m_{M}%
}G\left(  z_{2}\right) ;  \label{40}
\end{equation}
but, from (36) and (37), we can write for $M_{2}\gg m_{M}$
\begin{equation}
m_{\mu}^{\mbox{1-loop}}\simeq\frac{\xi_{2}^{2}}{128\pi^{2}}m_{M}F(z_{2});
\label{41}
\end{equation}
combining these equations, we obtain%
\begin{equation}
\Delta a_{\mu}^{FCh}\simeq\frac{m_{\mu}^{2}}{M_{2}^{2}}\times\frac
{z_{2}[(1-z_{2})^{2}-2z_{2}\left(  1-z_{2}\right)  +2z_{2}^{2}\ln z_{2}%
)}{2(1-z_{2})^{3}}, \label{42}%
\end{equation}
where the function $P(z_{2})=\frac{z_{2}[(1-z_{2})^{2}-2z_{2}\left(
1-z_{2}\right)  +2z_{2}^{2}\ln z_{2})}{2(1-z_{2})^{3}}$ is plotted in the
Fig.5. \ In this way, if the mass of the muon is of radiative origin we obtain
$\Delta a_{\mu}^{FCh}\sim m_{\mu}^{2}/M_{2}^{2}$ $.$ An analogous result was
obtained by Marciano using a toy model \cite{17} .%



\begin{figure}
\centering
\mbox{{\epsfig{figure=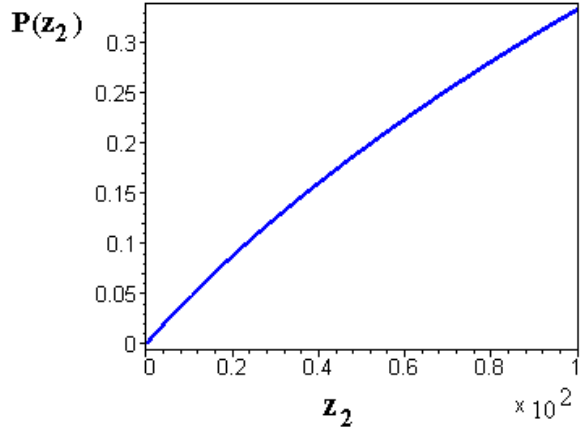,width=0.5\textwidth}}}
\vskip -0.5 cm
\caption{\label{fig5} Plot of $P(z_{2})$ . Note that $P(z_{2})$ is roughly $\mathcal{O(}
1\mathcal{)}$ on the values range considered.}
\end{figure}

\section{Weinberg symmetry invariance  }

In terms of the mixing angles $\theta_{L,R}$\bigskip, from the bi-unitary
diagonalization $U_{L}^{\dagger}\mathbf{M}U_{R}=\mathbf{M}_{\mathbf{diag}}$, we
find for the masses
\begin{eqnarray}
m_{\mu}  &  =\frac{1}{\sqrt{2}}[(c_{L}f_{0}-s_{L}f_{3})\upsilon_{1}%
c_{R}-(c_{L}f_{2}\upsilon_{3}-s_{L}f_{1}\upsilon_{2})s_{R}],\label{43}\\
m_{M}  &  =\frac{1}{\sqrt{2}}[(s_{L}f_{0}+c_{L}f_{3})\upsilon_{1}s_{R}%
+(s_{L}f_{2}\upsilon_{3}+c_{L}f_{1}\upsilon_{2})c_{R}], \label{44}%
\end{eqnarray}
where $\theta_{L,R}$  are given in (24) and (25), respectively. Under the
WS in (6): $\varphi\longrightarrow-\varphi$ , (equivalently
$\upsilon_{1}\longrightarrow-\upsilon_{1}\,$), $\theta_{L}$ is invariant, but
$\theta_{R}\longrightarrow-\theta_{R}$, then $m_{\mu}\longrightarrow-\,m_{\mu
}$ and $m_M $ is invariant $m_{M}\rightarrow m_{M},$. Now, let us remember
that $\mu$ and $M$ are in the same fundamental representation $\left[
\mathbf{27}\right]  $ of $E_{6}$. This entails that under WS invariance, we
will have $M_{L}\longrightarrow-M_{L}$ , $M_{R}\longrightarrow M_{R}.$ Then,
the mass eigenstates transform as:
\begin{eqnarray}
\mu_{L}^{0}  &  =c_{L}\mu_{L}-s_{L}M_{L}\longrightarrow-\mu_{L}^{0}%
\,,\nonumber\\
\mu_{R}^{0}  &  =c_{R}\mu_{R}-s_{R}M_{R}\ {\large \rightarrow\,}\mu_{R}%
^{0}\,,\nonumber\\
\widehat{M}_{L}  &  =s_{L}\mu_{L}+c_{L}M_{L}\longrightarrow-\widehat{M}%
_{L}\,,\nonumber\\
\widehat{M}_{R}  &  =s_{R}\mu_{R}+c_{R}M_{R}\,\,{\large \rightarrow
\,}\,\widehat{M}_{R}\,. \label{45}%
\end{eqnarray}
Thus, the WS invariance is ensured only when $\theta_{R}\longrightarrow0$ or
when $\upsilon_{2}\gg\upsilon_{1}.$ Consequently, the last\smallskip
\ transformations imply $m_{\mu}\rightarrow-m_{\mu}$, but not $m_{M}%
\,{\large \rightarrow}-m_{M}$ and then one may expect a linear correction
to\smallskip\ the muon magnetic moment as (31). This analysis do not apply
if the muon gets its mass by radiative corrections from other particles.

\section{\textbf{General Conclusions}}

To conclude, it is possible to explain the muon anomaly in our model based on
$E_{6}$ through  the breaking chain (\ref{1}), using only Higgses in $\mathbf{[78]}$
and $\mathbf{[351]}$ representations with a minimal set of Higgses to be
singlets and doublet under the SM symmetry. We find a linear relation between
masses for the muon anomaly ,if the radiative correction to muon mass, due to mixing
with heavy fermion, is small and WS is broken. On the other hand, we find a
quadratic relation between masses whenever we suppose that the muon has its
 mass generated only by radiative corrections in the GUT scale, since, in this case, WS
is conserved.

\vspace{.3 true cm}

\noindent
\textbf{Acknowledgements}

\vspace{.3 true cm}

\smallskip H. Ch. acknowledges the IF of the UFRJ and LAFEX-CBPF/MCT for
the kind hospitality; FAPERJ is acknowledged for his post-doctoral fellowship. 
C. N. F. and J. A. H.-N. express their gratitude to CNPq-Brazil for the 
invaluable financial help.

\end{document}